\newcommand{\diracslash}[1]{#1\llap{/\kern2pt}}

\newcommand{\be}{\begin{equation}}
\newcommand{\ee}{\end{equation}}
\newcommand{\bea}{\begin{eqnarray}}
\newcommand{\eea}{\end{eqnarray}}
\newcommand{\ba}[1]{\begin{array}{#1}}
\newcommand{\ea}{\end{array}}

\newcommand{\bt}{\begin{tabular}}
\newcommand{\et}{\end{tabular}}

\newcommand{\beas}{\begin{eqnarray*}}
\newcommand{\eeas}{\end{eqnarray*}}

\documentclass[preprint,prd,aps,amssymb,amsmath
,floats,nofootinbib,floatfix]{revtex4}

\DeclareSymbolFont{rsfs}{U}{rsfs}{m}{n}
\DeclareSymbolFontAlphabet{\mathrsfs}{rsfs}

\usepackage{graphicx}
\usepackage{multirow}
\usepackage{graphicx,epstopdf}

\usepackage{graphicx}
\usepackage{float}

\usepackage[utf8]{inputenc}
\usepackage[english]{babel}
\usepackage{hyperref}
\hypersetup{
    colorlinks=true,
    linkcolor=blue,
    filecolor=magenta,     
    citecolor=blue, 
    urlcolor=blue,
}
 
\usepackage{cleveref}

\begin{document}

\title{Charmonia and Bottomonia in asymmetric  magnetized hot nuclear matter} 

\author{Rajesh Kumar}
\email{rajesh.sism@gmail.com}
\affiliation{Department of Physics, Dr. B R Ambedkar National Institute of Technology Jalandhar, 
 Jalandhar -- 144011,Punjab, India}
\author{Arvind Kumar}
\email{iitd.arvind@gmail.com, kumara@nitj.ac.in}
\affiliation{Department of Physics, Dr. B R Ambedkar National Institute of Technology Jalandhar, 
 Jalandhar -- 144011,Punjab, India}

\def\be{\begin{equation}}
\def\ee{\end{equation}}
\def\bearr{\begin{eqnarray}}
\def\eearr{\end{eqnarray}}
\def\zbf#1{{\bf {#1}}}
\def\bfm#1{\mbox{\boldmath $#1$}}
\def\hf{\frac{1}{2}}
\def\kp{\zbf k+\frac{\zbf q}{2}}
\def\km{-\zbf k+\frac{\zbf q}{2}}
\def\hwo{\hat\omega_1}
\def\hwt{\hat\omega_2}

\begin{abstract}

We investigate the mass-shift of $P$-wave charmonium (${\chi_c}_0$, ${\chi_c}_1$)  and $S$ and $P$-wave  bottomonium ($\eta_b$, $\Upsilon$, ${\chi_b}_0$ and ${\chi_b}_1$) states in   magnetized hot asymmetric nuclear matter  using the unification of  QCD sum rules (QCDSR) and chiral $SU(3)$ model. Within QCDSR, we use two approaches, $i.e.$, moment sum rule and Borel sum rule. The magnetic field induced  scalar gluon condensate $\left\langle \frac{\alpha_{s}}{\pi} G^a_{\mu\nu} {G^a}^{\mu\nu} 
\right\rangle$ and  the twist-2 gluon operator $\left\langle  \frac{\alpha_{s}}{\pi} G^a_{\mu\sigma}
{{G^a}_\nu}^{\sigma} \right\rangle $  calculated in  chiral $SU(3$) model are utilised  in QCD sum rules to calculate the in-medium mass-shift of above mesons. The attractive mass-shift of these mesons is observed which is more sensitive to magnetic field in high density regime for charmonium, but less for bottomonium. These results  may be helpful to understand the decay of higher quarkonium states to the lower quarkonium  states in asymmetric heavy ion collision experiments.
\end{abstract}

\maketitle

\maketitle

\section{Introduction}
\label{intro}

To understand the confinement nature of QCD, the study of the spectral property of quarkonia in hot and dense asymmetric hadronic matter is very imperative. The in-medium properties of excited charmonium states $(\chi_c$ and $\psi$) and open charm mesons ($D$ and $\bar D$), can alter the $J/\psi$ production rate and its suppression in non-central heavy ion collisions \cite{Matsui1986,Song2009}. The suppression mechanism suggested by Matsui and Satz cannot fully explain the recent experimental data which shows nontrivial suppression \cite{Gavin1996,Gavin1997,Gavin1990}. Therefore,  to understand the additional suppression mechanism various other methods such as density dependent suppresion and comover scattering have been proposed \cite{Gavin1996,Cassing1997,Karsch2006}. In addition to temperature and density,  it is believed that in non-central  heavy ion collisions such as Large Hadron Collider (LHC) and Relativistic Heavy Ion Collider (RHIC) immense magnetic field is produced and are  approximated to be  the order of $eB$ $\sim$ $2-15 m_{\pi}^2$  (1${{m}_{\pi}^2}$ = $ 2.818\times 10^{18}$ gauss) \cite{Kharzeev2008,Fukushima2008,Skokov2009}. The decay of magnetic field does not occur immediately due to its interaction with the residual matter which is called chiral magnetic effect\cite{Kharzeev2013,Fukushima2008,Vilenkin1980,
Burnier2011}. According to Lenz's law, this magnetic field generates the induced currents in the remnant matter, which further produce the induced magnetic field that affects  the relaxation time and electric conductivity of the medium, hence the decay of external magnetic field\cite{Tuchin2011,Tuchin2011a,Tuchin2013,Marasinghe2011,Das2017,
Reddy2018,Cho2015}. The presence of magnetic field may cause the modification of chiral condensates \cite{Bali2013,Ozaki2014,Hong1998,Semenoff1999} and this may further have impact on the properties of charmonium and bottomonium states \cite{Cho2015,Cho2014,Reddy2018,Kumar2018,Jahan2018}. 
In the future experiments such as compressed baryonic matter (CBM) of FAIR collaboration, Japan Proton Accelerator Research Complex (J-PARC) and Nuclotron-based Ion Collider Facility (NICA,) significant observables are expected which may shed light on  the medium modifications of charmonia and bottomonia in the presence of density, temperature and magnetic field \cite{Rapp2010}.

In order to explain QCD phase diagram in non-perturbative region, various phenomenological models are proposed on the basis of effective field theory namely Walecka
model \cite{Walecka1974}, Nambu-Jona-Lasinio (NJL) model \cite{Nambu1961},  the Polyakov loop extended NJL  (PNJL) model \cite{Fukushima2004,Kashiwa2008,Ghosh2015},  Quark-Meson Coupling (QMC) model \cite{Guichon1988,Hong2001,Tsushima1999,Sibirtsev1999,Saito1994,Panda1997},  Polyakov Quark Meson (PQM) model \cite{Chatterjee2012,Schaefer2010},  coupled channel approach \cite{Tolos2004,Tolos2006,Tolos2008,Hofmann2005}, chiral $SU(3)$ model \cite{Papazoglou1999,Mishra2004a,Mishra2009,Kumar2010,
Kumar2018} and QCD sum rules  \cite{Reinders1981,Hayashigaki2000,
Hilger2009,Reinders1985,Klingl1997,Klingl1999}. These models incorporate the basic QCD properties such as trace anomaly and chiral symmetry.  The Polyakov extended models\cite{Skokov2010,Schaefer2007} and Functional Remormalization Group (FRG)\cite{Herbst2014,Drews2013} techniques incorporates the  thermal and quantum fluctuations near critical temperature region. Under these beyond mean field approximations, the critical point moves towards the larger chemical potential side\cite{Nakano2010,Herbst2011}.   In the previous work, we have studied the magnetic field induced in-medium properties of lowest charmonium states $J/\psi$ and $\eta_c$ using the QCDSR and shown that the mass-shift reflects the behaviour of gluon condensates which were incorporated from chiral $SU(3)$ model \cite{Kumar2018}. In the present paper, we intend to investigate the effect of external magnetic field on $P$-wave  charmonium as well as $S$ and $P$-wave bottomonium  states at finite temperature and density.  In this article,  non-perturbative
mean field approximation  is used, in which the contributions from the fluctuations  is neglected. In this approximation,  the contributions from pseudoscalar and axial vector meson is neglected as all meson fields are treated as classical fields \cite{Mishra2004a,Papazoglou1999}. The chiral $SU(3)$ model is used vastly to investigate the in-medium properties of light mesons in nuclear and hyperonic matter \cite{Kumar2010,Zschiesche2004,Mishra2004}. To study the in-medium properties of heavy $D$ and $B$ mesons, this model is also generalized to $SU(4)$ and $SU(5)$ respectively
 \cite{Mishra2004a,Mishra2009,Chhabra2017,Chhabra2017a}. In this model, the scalar and tensorial gluon condensates are extracted in terms of in-medium non-strange meson field $\sigma$, the strange scalar meson field $\zeta$, the scalar isovector meson field $\delta$, the vector meson field $\omega$, the vector-isovector meson field $\rho$, and  the dilaton field $\chi$\cite{Papazoglou1999}. On the other hand, QCDSR which is a non-perturbative approach to investigate the confining nature of strong interactions has also been used extensively to study the in-medium properties of heavy mesons  \cite{Reinders1981,Reinders1985,Klingl1997,Cho2015,Cho2014,
Kumar2010,Klingl1999,Morita2012,Kim2001,Suzuki2013,
Mishra2014}. The open charm and bottomonium mesons have also been studied under the influence of strong magnetic field \cite{Reddy2018,Gubler2016,Machado2014}.

The in-medium properties of charmonia and bottomonia have been studied extensively in the literature \cite{Kumar2010,Kumar2018,Andronic2008,Song2009,Suzuki2013,
Jahanb2018,Cho2014,Cho2015,Alford2013,Lee2003,Jahan2018,Morita2010a}. Due to the interaction of gluon condensates in the hadronic medium, the quarkonia  undergo significant modifications. The $P$-wave charmonia with modified mass and decay width decays into $J/\psi$ and hence affect the production ratio $N_{J/\psi}/N_{\chi_c}$ in nuclear matter which further affects the suppression mechanism \cite{Andronic2008}. The medium modifications of masses and decay width of ${\chi_c}_0$ and ${\chi_c}_1$ mesons in hot gluonic matter as a function of temperature has been studied using QCD sum rules \cite{Song2009}. In this article, the authors have investigated the larger mass-shift and width broadening for $P$-wave than the lower charmonium states such as $J/\psi$ and $\eta_c$ \cite{Kumar2010,Morita2012}. In Ref.\cite{Lee2003,Jahan2018}, using QCD second order stark effect, the in-medium properties of heavy charmonium states ($\psi$(3686) and $\psi$(3770)) has been studied in the nuclear matter  and observed a significant mass-shift. Also, in Ref. \cite{Cho2015}, authors have studied the effect of magnetic field on the lowest charmonium states using QCD sum rules. Magnetic field induced mass mixing of $J/\psi$ and $\eta_c$ has also been calculated under QCD sum rules approach \cite{Cho2014}. The mass-shift of bottomonia is studied by the unification of QCDSR and Maximum Entropy Method (MEM) in Ref.\cite{Suzuki2013} and calculted the bottomonia mass-spectra for different temperature. In Ref.\cite{Alford2013}, authors have investigated the mass-splitting by studying the mixing of pseudoscalar and vector charmonia and bottomina in the cornell potential due to the background magnetic field. The mass modifications of Upsilon 
$\Upsilon$ bottomonium state has been studied using the chiral $SU(3)$ model in the magentized nuclear matter at zero temperature and observed that the magnetic field effects are more prominent in high density regime \cite{Jahanb2018}. 

 The paper is organised as follows: In the next section, we will briefly describe the methodology to calculate the mass spectra of  charmonia and bottomonia. In section \ref{sec:4}, we discuss the quantitative results of the present work and finally section \ref{sec:5} is devoted to the conclusion.

\section{ Formalism}
\label{sec:2}

To study the effective mass of  charmonium and bottomonium, we use the conjunction of QCDSR with chiral $SU(3)$ model, which are the non-perturbative mechanism to study the confining nature of QCD \cite{Reinders1985,Reinders1981,Klingl1999,Mishra2004,Papazoglou1999}. 
 In subsection A, we briefly explain how gluon condensated are extracted from Chiral SU(3) model. Borel sum rule and moment sum rule are explained in subsection B.
\
 \subsection{Gluon Condensates from Chiral $SU(3)$ Model }
 \label{subsec:2.1}
 To describe nucleon-nucleon interactions, we use an effective field theoretical approach which is based on the non-linear realization of chiral symmetry \cite{Weinberg1968,Coleman1969,Bardeen1969} and  broken scale invariance                                              \cite{Papazoglou1999,Mishra2004,Mishra2004a} in the presence of external magnetic field at finite temperature  and density. These interactions are expressed in terms of the scalar fields $\sigma$, $\zeta$, $\delta$, $\chi$ and vector fields $\omega$ and $\rho$ in this model. The $\rho$ and $\delta$ fields are incorporated to introduce the effect of isospin asymmetry in nuclear medium. Also, the dilaton $\chi$ field is introduced to incorporate the trace anomaly property of QCD \cite{Papazoglou1999}. The Lagrangian density of chiral $SU(3)$ model under mean-field approximation  is given as
\be
{\cal L}_{chiral} = {\cal L}_{kin}  + {\cal L}_{vec} + \sum_{ M =S,V}{\cal L}_{MN}
        + {\cal L}_{SB}  + {\cal L}_0 .
\label{genlag} \ee

In above, ${\cal L}_{kin}$ represents the kinetic energy term,  $ {\cal L}_{vec}$ produces the mass of vector mesons through the interactions with scalar fields and carries the  self-interaction quartic terms, ${\cal L}_{MN}$ is the meson-nucleon interaction term, 
 where $S$ and $V$ represent the pseudoscalar and  vector mesons, respectively. The in-medium mass of nucleons is given as $m_{i}^{*} = -(g_{\sigma i}\sigma + g_{\zeta i}\zeta + g_{\delta i}\tau_3 \delta)$. Here, $g_{\sigma i}$, $g_{\zeta i}$ and $g_{\delta i}$ represent the coupling strengths of  nucleons ($i$=$n,p$) with  $\sigma$, $\zeta$ and $\delta$ fields respectively and $\tau_3$ is the third component of isospin.    ${\cal L}_{SB} $ describes the explicit chiral symmetry breaking, and $ {\cal L}_{0}$ describes  the spontaneous chiral symmetry breaking.

The thermopotential $\Omega$, per unit volume in zero magnetic field can be written as \cite{Zschiesche1997,Kumar2018} 

\begin{equation}
\label{thermo}
\frac{\Omega} {V}= -\frac{\gamma_i T}
{(2\pi)^3} \sum_{i = n\,, p\, }
\int d^3k\biggl\{{\rm ln}
\left( 1+e^{-\beta [ E^{\ast}_i(k) - \mu^{*}_{i}]}\right) \\
+ {\rm ln}\left( 1+e^{-\beta [ E^{\ast}_i(k)+\mu^{*}_{i} ]}
\right) \biggr\} -{\cal L}_{vec} - {\cal L}_{SB}- {\cal L}_0 -{\mathcal{V}}_{vac},   
\end{equation}
where the sum runs over neutron and proton, $\gamma_i$ is 
spin degeneracy factor for nucleons, $ \mu^{*}_{i}=\mu_{i}-g_{\omega i}\omega-g_{\rho i}\tau_{3}\rho$ and 
$E^{\ast}_i(k)=\sqrt{k^2+{m^{*}_{i}}^2}$ 
 are the effective nucleon chemical potential, and  effective single particle
energy of nucleons,  respectively. Also, vacuum potential energy, ${\mathcal{V}}_{vac}$  is subtracted from the thermopotential in order to get zero vacuum energy.

 The Lagrangian density (Eq.(\ref{genlag})) in the presence of magnetic field modifies as

\begin{equation}
{\cal L}={\cal L}_{chiral}+{\cal L}_{mag},
\label{Tlag}
\end{equation}
where
\be 
{\cal L}_{mag}=-{\bar {\phi_i}}q_i 
\gamma_\mu A^\mu \phi_i
-\frac {1}{4} \kappa_i \mu_N {\bar {\psi_i}} \sigma ^{\mu \nu}F^{\mu \nu}
\phi_i
-\frac{1}{4} F^{\mu \nu} F_{\mu \nu}.
\label{lmag}
\ee

In above $\phi_i$ is a wave function of $i^{th}$ nucleon, and the next term represents the interaction
with the electromagnetic field tensor, $F_{\mu \nu}$. Further, $\mu_N$ and $k_i$  are the nuclear magneton and the anomalous magnetic moment of  $i^{th}$ nucleon, respectively.
In this article, the magnetic field is chosen to be uniform and along the
$Z$-axis. We will discuss the interactions of uncharged and charged particle with uniform magnetic field in the following. 

In the presence of uniform external magnetic field, the charged proton experiences Lorentz force, hence Landau quantization comes into picture. In this quantization, the transverse momenta of
proton are confined to the closed loop namely Landau levels, $\nu$, with, $k_\perp^2 = 2 \nu |q_p| B$, where 
$\nu \geq 0$ is a quantum number \cite{Strickland2012}. Thus, the integral transformation takes place as 
$\int {d^3}k \rightarrow \frac{|q_p| B}{(2\pi)^2} 
\sum_{\nu} \int_{0}^\infty d k_\parallel \, ,$
where $k_\parallel $ is the longitudinal momenta. Here, the summation represents a sum over the discrete Landau levels, $\nu=n+\frac{1}{2}-\frac{q_p}{|q_p|}\frac{s}{2}=0, 1, 2, \ldots$ of proton in the normal plane. 
Also, $n$ is the orbital quantum number and the quantum 
number $s$ is $-1$ for spin down and $+1$ for spin up protons. The effective single particle energy of proton  also gets modified  as $\tilde E^{p}_{\nu, s}=\sqrt{\left(k^{p}_{\parallel}\right)^{2}+
\left(\sqrt{m^{* 2}_{p}+2\nu |q_{p}|B}-s\mu_{N}\kappa_{p}B \right)^{2}}$\cite{Ternov1966}.

Under these modifications, the first term of thermopotential (Eq.(\ref{thermo})), for proton will be
\bea
\label{thermop}
&\frac{\Omega_p} {V}= -
\frac{ T|q_p| B}{(2\pi)^2} \Bigg[
\sum_{\nu=0}^{\nu_{max}^{(s=1)}} \int_{0}^\infty d k_\parallel \, \biggl\{{\rm ln}
\left( 1+e^{-\beta [ \tilde E^{p}_{\nu, s} - \mu^{*}_{p} ]}\right) 
+ {\rm ln}\left( 1+e^{-\beta [ \tilde E^{p}_{\nu, s}+\mu^{*}_{p} ]}
\right) \biggr\}\nonumber\\
&+
\sum_{\nu=1}^{\nu_{max}^{(s=-1)}} \int_{0}^\infty d k_\parallel \, \biggl\{{\rm ln}
\left( 1+e^{-\beta [ \tilde E^{p}_{\nu, s} - \mu^{*}_{p} ]}\right)+ {\rm ln}\left( 1+e^{-\beta [ \tilde E^{p}_{\nu, s}+\mu^{*}_{p} ]}
\right) \biggr\}\Bigg].  
\eea

On the other hand, there is no Landau quantization for an uncharged neutron in the presence of external magnetic field, therefore the integral $\int  \frac{d^3k}{(2\pi)^3} \,$ remains unmodified \cite{Strickland2012}. But due to non-zero anomalous magnetic moment $k_n$, the effective single particle energy of neutron gets modified as $\tilde E^{n}_{s}= \sqrt{\left(k^{n}_{\parallel}\right)^{2} +
\left(\sqrt{m^{* 2}_{n}+\left(k^{n}_{\bot}\right)^{2} }-s\mu_{N}\kappa_{n}B 
\right)^{2}}$ and hence, the first term of Eq.(\ref{thermo}) for neutron, will be

\begin{equation}
\label{thermon}
\frac{\Omega_n} {V}= -\frac{T}
{(2\pi)^3} \sum_{s=\pm 1}
\int d^3k\biggl\{{\rm ln}
\left( 1+e^{-\beta [ \tilde E^{n}_{s} - \mu^{*}_{n} ]}\right) \\
+ {\rm ln}\left( 1+e^{-\beta [\tilde E^{n}_{s}+\mu^{*}_{n}]}
\right) \biggr\}.
\end{equation}

The net thermopotential in the presence of uniform external magnetic field can be expressed as

\begin{equation}
\label{thermonet}
\frac{\Omega} {V}= \frac{\Omega_p} {V}+\frac{\Omega_n} {V} -{\cal L}_{vec}- {\cal L}_{SB} - {\cal L}_0 -{\mathcal{V}}_{vac}. 
\end{equation}

By minimizing the
thermopotential $\Omega/V$ with respect to the  fields $\sigma$, $\zeta$, $\delta$, $\chi$, $\rho$ and $\omega$, the corresponding coupled equations of motion of these fields  are determined and given as

\begin{eqnarray}
&&\frac{\partial (\Omega/V)}{\partial \sigma}= k_{0}\chi^{2}\sigma-4k_{1}\left( \sigma^{2}+\zeta^{2}
+\delta^{2}\right)\sigma-2k_{2}\left( \sigma^{3}+3\sigma\delta^{2}\right)
-2k_{3}\chi\sigma\zeta \nonumber\\
&-&\frac{d}{3} \chi^{4} \bigg (\frac{2\sigma}{\sigma^{2}-\delta^{2}}\bigg )
+\left( \frac{\chi}{\chi_{0}}\right) ^{2}m_{\pi}^{2}f_{\pi}
-\sum g_{\sigma i}\rho_{i}^{s} = 0,
\label{sigma}
\end{eqnarray}
\begin{eqnarray}
&&\frac{\partial (\Omega/V)}{\partial \zeta}= k_{0}\chi^{2}\zeta-4k_{1}\left( \sigma^{2}+\zeta^{2}+\delta^{2}\right)
\zeta-4k_{2}\zeta^{3}-k_{3}\chi\left( \sigma^{2}-\delta^{2}\right)\nonumber\\
&-&\frac{d}{3}\frac{\chi^{4}}{\zeta}+\left(\frac{\chi}{\chi_{0}} \right)
^{2}\left[ \sqrt{2}m_{K}^{2}f_{K}-\frac{1}{\sqrt{2}} m_{\pi}^{2}f_{\pi}\right]
 -\sum g_{\zeta i}\rho_{i}^{s} = 0 ,
\label{zeta}
\end{eqnarray}
\begin{eqnarray}
&&\frac{\partial (\Omega/V)}{\partial \delta}=k_{0}\chi^{2}\delta-4k_{1}\left( \sigma^{2}+\zeta^{2}+\delta^{2}\right)
\delta-2k_{2}\left( \delta^{3}+3\sigma^{2}\delta\right) +2k_{3}\chi\delta
\zeta \nonumber\\
& + &  \frac{2}{3} d \chi^4 \left( \frac{\delta}{\sigma^{2}-\delta^{2}}\right)
-\sum g_{\delta i}\tau_3\rho_{i}^{s} = 0 ,
\label{delta}
\end{eqnarray}

\begin{eqnarray}
\frac{\partial (\Omega/V)}{\partial \omega}=\left (\frac{\chi}{\chi_{0}}\right) ^{2}m_{\omega}^{2}\omega+g_{4}\left(4{\omega}^{3}+12{\rho}^2{\omega}\right)-\sum g_{\omega i}\rho_{i}^{v} = 0 ,
\label{omega}
\end{eqnarray}

\begin{eqnarray}
\frac{\partial (\Omega/V)}{\partial \rho}=\left (\frac{\chi}{\chi_{0}}\right) ^{2}m_{\rho}^{2}\rho+g_{4}\left(4{\rho}^{3}+12{\omega}^2{\rho}\right)-\sum g_{\rho i}\tau_3\rho_{i}^{v} = 0 ,
\label{rho}
\end{eqnarray}
and
\begin{eqnarray}
&&\frac{\partial (\Omega/V)}{\partial \chi}=k_{0}\chi \left( \sigma^{2}+\zeta^{2}+\delta^{2}\right)-k_{3}
\left( \sigma^{2}-\delta^{2}\right)\zeta + \chi^{3}\left[1
+{\rm {ln}}\left( \frac{\chi^{4}}{\chi_{0}^{4}}\right)  \right]
+(4k_{4}-d)\chi^{3}
\nonumber\\
&-&\frac{4}{3} d \chi^{3} {\rm {ln}} \Bigg ( \bigg (\frac{\left( \sigma^{2}
-\delta^{2}\right) \zeta}{\sigma_{0}^{2}\zeta_{0}} \bigg )
\bigg (\frac{\chi}{\chi_0}\bigg)^3 \Bigg )+
\frac{2\chi}{\chi_{0}^{2}}\left[ m_{\pi}^{2}
f_{\pi}\sigma +\left(\sqrt{2}m_{K}^{2}f_{K}-\frac{1}{\sqrt{2}}
m_{\pi}^{2}f_{\pi} \right) \zeta\right] \nonumber\\
&-& \frac{\chi}{{{\chi_0}^2}}(m_{\omega}^{2} \omega^2+m_{\rho}^{2}\rho^2)  = 0 ,
\label{chi}
\end{eqnarray}

respectively.

In above,  $f_K$, $f_\pi$ and $m_K$, $m_\pi$  are the decay constants and masses of $K$, $\pi$ mesons, respectively and the additional parameters $k_0, k_2$ and $k_4$ are fitted to reproduce the vacuum values of scalar $\sigma$, $\zeta$ and $\chi$ mesons and the remaining parameters,  $k_1$ is selected to generate the effective mass of nucleon at nuclear saturation density (around 0.$65 m_N$) and $k_3$ is constrained by $\eta^\prime$  and $\eta$  masses. Furthermore, $\rho^{v}_{i}$ and $\rho^{s}_{i}$ represent the number/vector and scalar densities of $i^{th}$ nucleon respectively for magnetic/non-magnetic case \cite{Kumar2018,Rabhi2011,Broderick2000,Broderick2002}. Also, the isospin asymmetry of the medium is defined through parameter $\eta=(\rho^{v}_{n} -{\rho^{v}_{p}} )/2\rho_N$.

%

Also, the scalar gluon condensate $G_0$=$\left\langle \frac{\alpha_{s}}{\pi} 
G^a_{\mu \nu} {G^a}^{\mu \nu} \right\rangle $ , and tensorial gluon operator $G_2$=$\left\langle \frac{\alpha_{s}}{\pi} G^a_{\mu\sigma} {{G^a}_\nu}^{\sigma} \right\rangle $  can be expressed within chiral $SU$(3) model in terms of  scalar fields through following relation\cite{Kumar2010}

\begin{eqnarray}
G_0=  \frac{8}{9} \Bigg [(1 - d) \chi^{4}
+\left( \frac {\chi}{\chi_{0}}\right)^{2} 
\left( m_{\pi}^{2} f_{\pi} \sigma
+ \big( \sqrt {2} m_{K}^{2}f_{K} - \frac {1}{\sqrt {2}} 
m_{\pi}^{2} f_{\pi} \big) \zeta \right) \Bigg ],
\label{chiglum}
\end{eqnarray}

and

\begin{equation}
G_2= \frac{\alpha_{s}}{\pi}\Bigg [-(1-d+4 k_4)(\chi^4-{\chi_0}^4)-\chi ^4 {\rm {ln}}
\Big (\frac{\chi^4}{{\chi_0}^4}\Big )
 +  \frac {4}{3} d\chi^{4} {\rm {ln}} \Bigg (\bigg( \frac {\left( \sigma^{2} 
- \delta^{2}\right) \zeta }{\sigma_{0}^{2} \zeta_{0}} \bigg) 
\bigg (\frac {\chi}{\chi_0}\bigg)^3 \Bigg ) \Bigg ]
\label{g2approx}
\end{equation}
respectively.
The value of $d$  has been taken from QCD beta function, $\beta_{QCD}$  at the one loop level \cite{Papazoglou1999}, 
with $N_f$ flavors  and $N_c$ colors, 
\be
\label{qcdbeta}
   \beta_{QCD}=-\frac{11 N_c g^3}{48 \pi^2} \left(1-\frac{2N_f}{11 N_c}\right)
   +{\cal O}(g^5) . 
\ee                               

In the limit, quark mass term (second term) equals to zero, the scalar gluon condensate given by Eq.(\ref{chiglum}) gets modified as
\begin{eqnarray}
G_0=  \frac{8}{9}  (1 - d) \chi^{4}
 .
\label{chiglu}
\end{eqnarray}

\subsection{ In-medium Mass of Heavy Quarkonia from QCD Sum Rule}
\label{subsec:2.2}
In QCDSR, we use two approaches, $i.e.$,   Borel sum rules and moment sum rules. The moment sum rules are derived using the operator product expansion(OPE) method, which is a non perturbative technique to define the product of fields as a sum over the same fields to resolve the difficulties arising from perturbative effects\cite{Reinders1981}. Whereas, Borel sum rules are derived from Borel transformations and OPE. These transformations are standard mathematical technique to incorporate perturbative effects by summation  of divergent asymptotic series\cite{Morita2010a}. The moment sum rule is a good tool to calculate heavy quark masses due to large separation scale, nevertheless, Borel sum rules has several merits, for example better OPE convergence.  

The current-current correlation function of heavy quark currents is given by 
 \cite{Reinders1985}
\begin{equation}
\Pi^{\, J}(p) = i \int d^4x e^{i p \cdot x} \langle T[j^{\, c}(x) j^{\, c \dag}(0)] \rangle , \label{correlation function}
\end{equation}
with $\left\langle  \right\rangle$ being the Gibbs average. In above, $p=( E_0, \vec{p})$, is four-momentum vector and symbol  $c$ represents the scalar $(S)$, vector $(V)$, pseudoscalar ($P$) and  axial vector $(A)$ mesons. These mesons currents are defined as $j^{\, S}=\bar{q}q$, $ j^{\, V}_\mu = \bar{q} \gamma_\mu q$, $j^{\, P}=\bar{q} \gamma_5 q$ and $ j^{\, A}_\mu = (q_\mu q_\nu/q^2-g_{\mu \nu})\bar{q} \gamma^\nu \gamma_5 q$ with $q$ being the heavy quark operator.

\subsubsection{ Borel sum rule}
\label{subsec:2.1.1}

 We perform Borel transformation \cite{Morita2010a} on the function $\Pi^{\, J}(p)$ by going in deep Euclidean region ($p^2$=$E^2_0$=-$\Omega^2\ll 0$) to obtain better radius of convergence. The transformed correlation function $\mathcal{M}^J(M^2)$  can be written as 
\begin{equation}
 \mathcal{M}^J(M^2)  =\lim_{\substack{\Omega^2/n \rightarrow M^2, \\ n,\Omega^2
  \rightarrow \infty}}
  \frac{(\Omega^2)^{n+1}\pi}{n!}\left(-\frac{d}{d\Omega^2}\right)^n
  \tilde{\Pi}^J(\Omega^2).
  \label{Borel_moment}
\end{equation}

At finite temperature and density of the medium, the correlation function can be expanded in term of dimensio-4 scalar gluon condensate and twist-2 gluon operator, whose Borel transformation, using Eq.(\ref{Borel_moment}), will lead to

\begin{equation}
 \mathcal{M}^J(M^2) = e^{-g}\pi
  A^J(g)[1+\alpha_s(M^2)a^J(g)+b^J(g)\phi_b +c^J(g)\phi_c],
 \label{eq:Borel_moment}
\end{equation}
In above $g=4m_q^2/M^2$ is a dimensionless scale parameter, where $M$ is Borel Mass. Symbols
$A^J(g)$, $a^J(g)$, $b^J(g)$ and $c^J(g)$ are Borel transformed Wilson coefficients and are  given in Ref.~\cite{Morita2010a}. The first term is the leading order of OPE. The second term corresponds to the perturbative correction and rest two terms are associated with the medium modified  scalar gluon condensate $G_0$=$\left\langle \frac{\alpha_{s}}{\pi} 
G^a_{\mu \nu} {G^a}^{\mu \nu} \right\rangle $ and tensorial gluon condensate  $G_2$=$\left\langle \frac{\alpha_{s}}{\pi} G^a_{\mu\sigma} {{G^a}_\nu}^{\sigma} \right\rangle $ as  

\begin{equation}
\phi_{b} = \frac{4 \pi^{2}}{9} \frac{G_0 }{(4 m_{q}^{2})^{2}},
\label{phib} 
\end{equation}
 
 and

\begin{equation}
\phi_{c} =  \frac{4 \pi^{2}}{3(4 m_{q}^{2})^{2}}G_2 ,
\label{phic}
\end{equation}
respectively, which are also the carrier of thermal effects of the medium. The  $\alpha_{s}$ and $m_{q}$  parameters are the  running coupling constant and running quark mass,
respectively. The consdensate $G_0$ and $G_2$ appearing in above equations are evaluated using Eqs.(\ref{chiglum}) and (\ref{g2approx}), respectively.

In QCDSR analysis, dispersion integral is used to express correlation function in terms of spectral function. For finite temperature and density of the medium, spectral function will be imaginary part of the correlation function\cite{Morita2010a}. Thus, we have following dispersion relation for Borel sum rule,

\begin{equation}
 \mathcal{M}^{\text{$J$}}(M^2) =
  \int_{0}^{\infty}ds\,e^{-s/M^2}\text{Im}\tilde{\Pi}^J(s). \label{eq:dispersion}
\end{equation}
In above, $e^{-s/M^2}$ is a exponential weight factor and $s$ represents a continuum parameter. 
%

In Eq.(\ref{eq:dispersion}),  Im$\tilde{\Pi}^J(s)$ can be decomposed into pole and the continuum part as
\begin{equation}
 \text{Im}\tilde{\Pi}(s) = \text{Im}\tilde{\Pi}^{\text{$pole$}}(s) + \text{Im}\tilde{\Pi}^{\text{$cont$}}(s),\label{eq:phen}
\end{equation}
with pole contribution under zero decay width approximation
\begin{align}
 \text{Im}\tilde{\Pi}^{\text{$pole$}}(s)= 
				        f_0\delta(s-{{m^*_{Q}}}^{2}),
				       \quad \Gamma=0,
				       \label{eq:phen_pole}
\end{align}
			where $f_0$ is a strength parameter. Finite decay width is important in deconfined medium, whereas in the present work, we investigated the modification of masses of charmonium and bottomonium in nuclear medium, which is confined in nature\cite{Morita2010a}. In Ref.\cite{Zschocke2002}, using the zero width approximation, authors have studied the mass of light vector mesons $\rho$, $\omega$ and $\phi$ under finite temperature and density.
The continuum contribution in Eq.(\ref{eq:phen}) is defined in terms of a perturbative spectral function with a sharp threshold factor, $i.e.$,       
\begin{align}
 \text{Im}\tilde{\Pi}^{\text{$cont$}}(s)=
 \theta(s-s_0)\text{Im}\tilde{\Pi}^{J,\text{pert}}(s),
 \label{eq:phen_cont}
\end{align}

where $\theta(s-s_0)$, is sharp threshold factor for the continuum, and the cut off parameter $s_0$ for the pole term  is adjusted to reproduce the corresponding side of OPE and $\tilde{\Pi}^{J,\text{pert}}(s)$ perturbative spectral function ,  for different meson currents\cite{Morita2010a}. Using Eq.(\ref{eq:phen}) in (\ref{eq:dispersion}), we have

\begin{eqnarray}
 \mathcal{M}^{\text{$J$}}(M^2) =
  \int_{0}^{s_0}ds\,e^{-s/M^2}\text{Im}\tilde{\Pi}^{pole}(s)+ \int_{0}^{\infty}ds\,e^{-s/M^2}\text{Im}\tilde{\Pi}^{cont}(s)
   \nonumber\\
\Rightarrow \mathcal{M}^{\text{$J$}}(M^2) =\int_{0}^{s_0}ds\,e^{-s/M^2}\text{Im}\tilde{\Pi}^{pole}(s)+\mathcal{M}^{\text{$cont$}}(M^2)
      \nonumber\\
\Rightarrow  \mathcal{M}^{\text{$J$}}(M^2) - \mathcal{M}^{\text{$cont$}}(M^2)= \int_{0}^{s_0}ds\,e^{-s/M^2}\text{Im}\tilde{\Pi}^{pole}(s).
\label{eq:spectralm}
\end{eqnarray}

Differentiating both sides of above equation with respect to $1/M^2$ and taking ratio with Eq.(\ref{eq:spectralm}) itself (to eliminate $f_0$ from the pole term from Eq.(\ref{eq:phen_pole})), we get

 \begin{gather}
  -\frac{\displaystyle
   \frac{\partial}{\partial(1/M^2)}
   [\mathcal{M}^{\text{$J$}}(M^2) -\mathcal{M}^{\text{$cont$}}(M^2)]}
   {\mathcal{M}^{\text{$J$}}(M^2) -\mathcal{M}^{\text{$cont$}}(M^2)} 
   = \frac{ \displaystyle \int_{0}^{s_0} ds \, s\,e^{-s/M^2}
   \text{Im}\tilde{\Pi}^{\text{$pole$}}(s)}
   {\displaystyle \int_{0}^{s_0} ds \, e^{-s/M^2}
   \text{Im}\tilde{\Pi}^{\text{$pole$}}(s)}={{m^*_{Q}}}^{2},
   \label{eq:sumrule}
 \end{gather}
 
 where ${{m^*_{Q}}}^{2}$ is because of use of Eq.(\ref{eq:phen_pole}) in 2$^{nd}$ term of Eq.(\ref{eq:sumrule}). We have solved above equation to evaluate the in-medium mass ${{m^*_{Q}}}$, of a given quarkonia as a function of Borel parameter $M^2$.

\subsubsection{ Moment sum rule}
\label{subsec:2.1.2}
 The  $n^{th}$ moment $M_n^J$, corresponds from the derivative of the correlation function, if one does not take the limit in Eq.(\ref{Borel_moment}), $i.e.$,

\begin{eqnarray}
\label{dispersion}
M_n^J &\equiv& \left. { 1 \over n!} \left( {d \over d {E_0}^2} \right)^n
\tilde{\Pi}^{J}({E_0}^2) \right|_{{E_0}^2=-Q^2}, \nonumber \\ &&=
\frac{1}{\pi}\int_{4m_{q}^2}^{\infty}\frac{\mbox{Im}
\tilde{\Pi}^{J}(s)}{(s+Q^2)^{n+1}}ds,
\end{eqnarray}
at a fixed $Q^2=4m_c^2 \xi$.

Within QCDSR, using OPE the moment can be expressed analytically as follows \cite{Klingl1997}
\begin{equation}
M_{n}^{J} (\xi) = A_{n}^{J} (\xi) \left[  1 + a_{n}^{J} (\xi) \alpha_{s} 
+ b_{n}^{J} (\xi) \phi_{b} + c_{n}^{J} (\xi) \phi_{c} \right],
\label{moment}  
\end{equation}
where $A_n^J(\xi)$, $a_n^J(\xi)$, $b_n^J (\xi)$ and $c_n^J (\xi)$
are the Wilson coefficients for meson currents\cite{Reinders1981,Song2009} and $\xi$ is the 
normalization scale. 
Hence by using $\phi_b$ and $\phi_c$ (Eqs.(\ref{phib}) and (\ref{phic})) in the expression of  $M_{n}^{J}$,  the in-medium mass of the quarkonium states\cite{Reinders1985,Kumar2010,Kumar2018} can be written as
\begin{equation}
{{m^*_{Q}}}^{2} \simeq \frac{M_{n-1}^{J} (\xi)}{M_{n}^{J} (\xi)} - 4 m_{q}^{2} \xi.
\label{masscharm}
\end{equation}

 The in-medium mass-shift of quarkonium is given by 
\begin{equation}
\Delta {{m^*_{Q}}}={{m^*_{Q}}}-{{m_{Q}}},
\label{mass-shift}
\end{equation}
where ${{m_{Q}}}$ ($Q={\chi_c}_0(S)$, ${\chi_c}_1(A),\eta_b(P), \Upsilon(V), {\chi_b}_0(S),{\chi_b}_1(A)$) is the vacuum mass of mesons. In the next section, we shall discuss the numerical results of present work.

\section{Numerical Results and Discussions}
\label{sec:4}
The in-medium mass-shift of charmonia and bottomonia is calculated through two different QCDSR approach by using in-medium scalar gluon condensate $G_0$, and tensorial gluon condensate $G_2$, in the presence of uniform external magnetic field at finite temperature and density of asymmetric nuclear matter. We have used the value of running charm(bottom) quark mass $m_c(m_b)$, and running strong coupling constant $\alpha_s$  to be $1240$ $(4120)$ MeV and $0.21(0.158)$, respectively. 
 Different parameters used in this work are listed in table \ref{ccc}.  We have divided this discussion in two subsection. In subsection A, we will discuss the in-medium behaviour of scalar fields $\sigma$, $\zeta$, $\delta$ and $\chi$ as a function of external magnetic field $eB$. The magnetic field induced  gluon condensates and mass-shift of heavy quarkonia will be discussed in subsection B.

\subsection{  Scalar Fields $\sigma$, $\zeta$, $\delta$, and $\chi$ in  Magnetized Hot Nuclear Matter} 
\label{fields}

\begin{figure}[h]
\includegraphics[width=16cm,height=16cm]{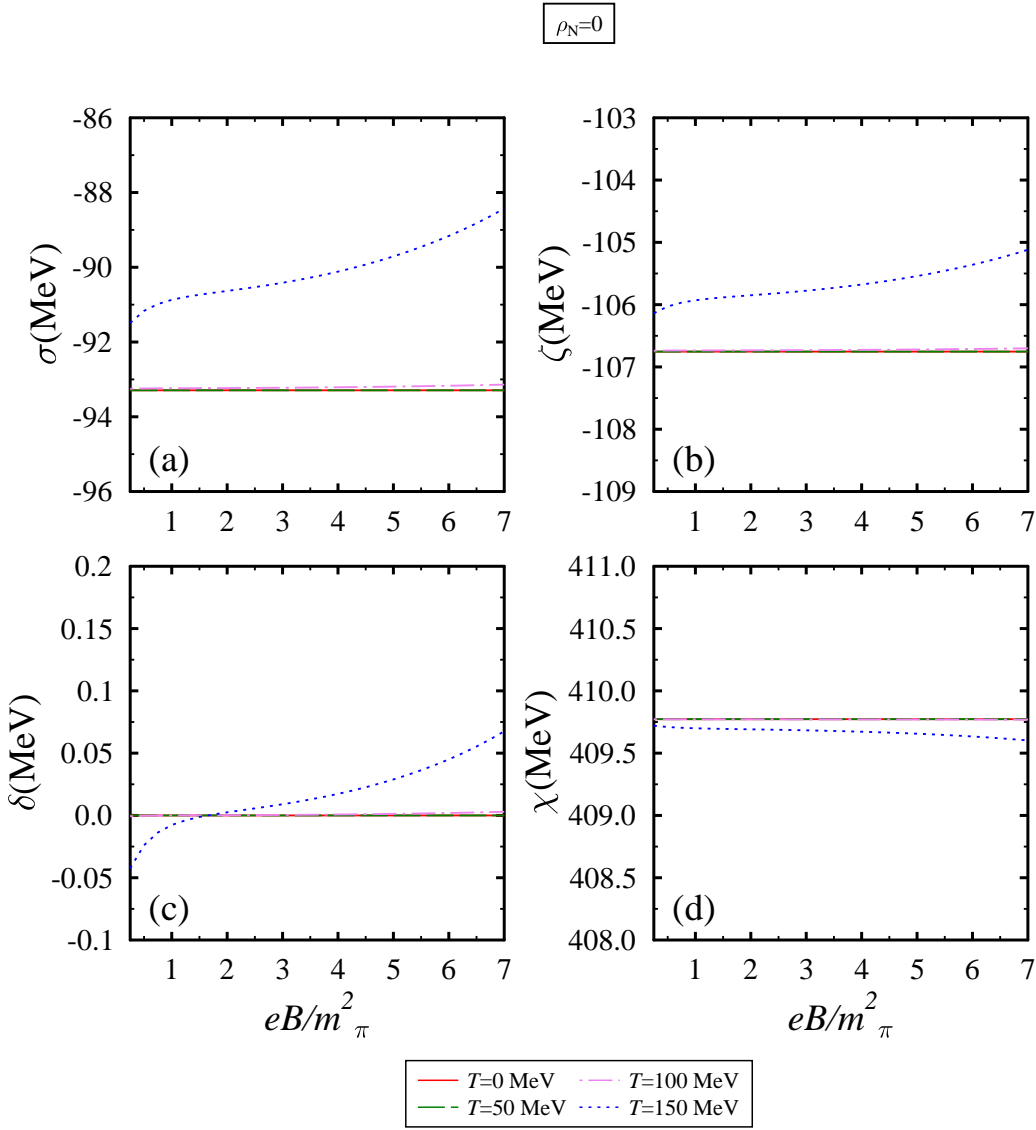}
\caption{(Color online) The scalar fields $\sigma$, $\zeta$, $\delta$ and $\chi$ plotted as a function of magnetic field $eB/{{m^2_{\pi}}}$, for different values of temperature $T$, at nucleon density $\rho_N=0$.}
\label{ffields0}
\end{figure}

In the chiral $SU(3)$ model, the effect of magnetic field comes into picture through the scalar density $\rho^{s}_i$, and vector density $\rho^{v}_i$, of the nucleons appearing in the scalar and vector field equations (\ref{sigma}) to (\ref{omega})\cite{Kumar2018}. 
 \begin{table}
\begin{tabular}{|c|c|c|c|c|}

\hline 
$\sigma_0$ (MeV) & $\zeta_0$ (MeV) & $\chi_0$ (MeV) & $d$ & $\rho_0$ ($\text{fm}^{-3}$)  \\ 
\hline 
-93.29 & -106.8 & 409.8 & 0.064 & 0.15  \\ 
\hline
\hline

\hline
$k_0$ & $k_1$ & $k_2$ & $k_3$ & $k_4$  \\ 
\hline 
2.53 & 1.35 & -4.77 & -2.77 & -0.218  \\ 
\hline
\hline

\hline 
$m_\pi $(MeV) &$ m_K$ (MeV) &$ f_\pi$ (MeV) & $f_K$(MeV) & $g_4$ \\ 
\hline 
139 & 498 & 93.29 & 122.14 & 79.91  \\ 
\hline
\hline

\hline
$g_{\sigma N}$  & $g_{\zeta N }$  &  $g_{\delta N }$  &
$g_{\omega N}$ & $g_{\rho N}$ \\

\hline 
10.56 & -0.46 & 2.48 & 13.35 & 5.48  \\ 
\hline
\hline

\end{tabular}
\caption{Values of various parameters.} \label{ccc}
\end{table} 
 In Fig.\ref{ffields0}, we have plotted the variation of
 $\sigma$, $\zeta$, $\delta$ and  $\chi$ fields as a function of magnetic field $eB/{{m^2_{\pi}}}$ (1${{m^2_{\pi}}}$ = $ 5.48\times 10^{-2}$ GeV$^2$=$ 2.818\times 10^{18}$ gauss), at nucleon density $\rho_N$=0 and temperatures $T$=0, 50, 100, 150 MeV. Here, $e$ and ${{m_{\pi}}}$ are the electric charge and mass of pion, respectively. As can be seen from this figure, the magnetic field affects the scalar fields negligibly for temperatures $T$=0, 50 and 100 MeV. But for $T$=150 MeV, the magnitude of $\sigma$, $\zeta$, and $\chi$  field decreases as a function of magnetic field. This is due to the fact that at higher temperature chiral symmetry restoration comes into picture and the different magnetic field strength rectifies the critical temperature which is in accordance with (inverse) magnetic catalysis process \cite{Kharzeev2013}. 

\begin{figure}
\includegraphics[width=16cm,height=21cm]{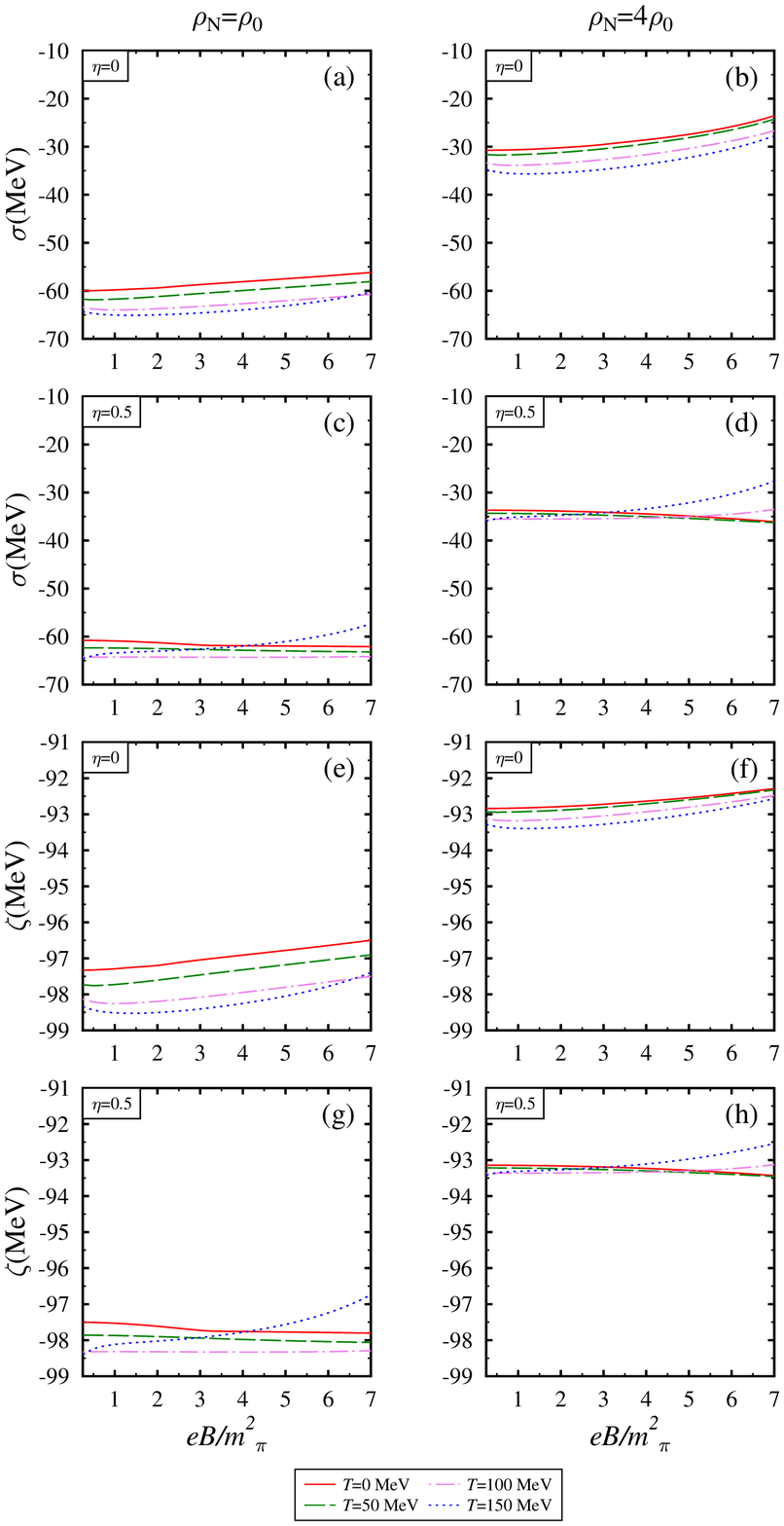}
\caption{(Color online) The scalar field $\sigma$ and $\zeta$ plotted as a function of magnetic field $eB/{{m^2_{\pi}}}$, for different values of temperature $T$, nucleon density $\rho_N$ and isospin asymmetry parameter, $\eta$.}
\label{fsigmazeta}
\end{figure}

\begin{figure}
\includegraphics[width=16cm,height=21cm]{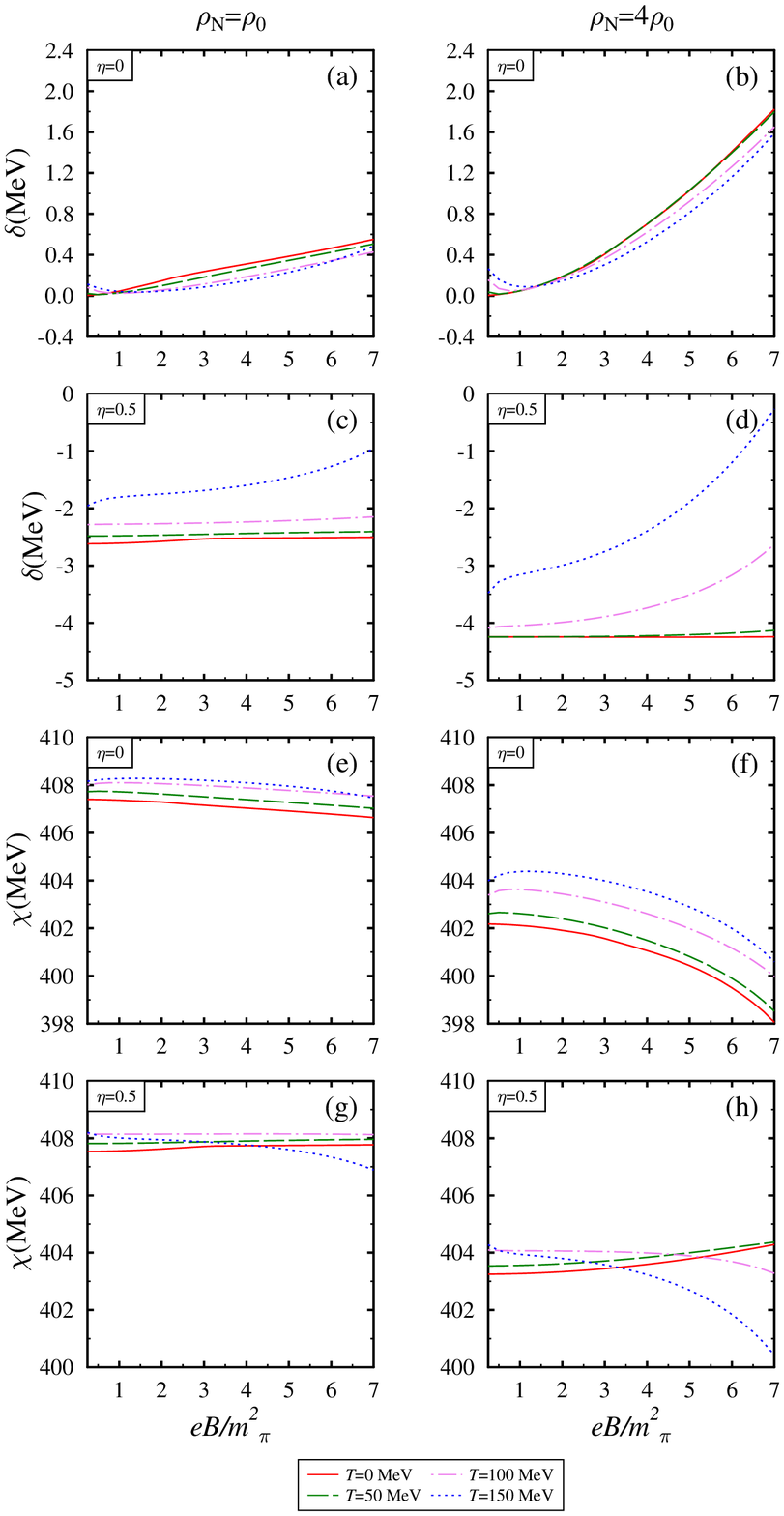}
\caption{(Color online) The scalar field $\delta$ and $\chi$ plotted as a function of magnetic field $eB/{{m^2_{\pi}}}$, for different values of temperature $T$, nucleon density $\rho_N$ and isospin asymmetry parameter, $\eta$.}
\label{fdeltachi}
\end{figure}

In Figs.\ref{fsigmazeta} and \ref{fdeltachi}, we have plotted the scalar fields  $\sigma$, $\zeta$, $\delta$ and  $\chi$ as a function of magnetic field $eB/{{m^2_{\pi}}}$ at different temperature $T$, for nucleon density $\rho_N$=$\rho_0$ ($4\rho_0$) and isospin asymmetry parameters $\eta$=0 and 0.5. In Fig.\ref{fsigmazeta}, for symmetric nuclear matter ($\eta=0$), it is observed that the magnitude of $\sigma$ and $\zeta$ fields decreases with the increase in magnetic field. The drop  is more in case of high density as compared to low density. For example, the values of $\sigma$ field at $T$=0 and  $\rho_N$=$\rho_0$ ($4\rho_0$)  are   -59.83 (-30.62) and -56.19 (-23.55) MeV for  $eB$=${{m^2_{\pi}}}$ and 7${{m^2_{\pi}}}$, respectively and the values of $\zeta$ fields are -97.29 (-96.50) and -92.83 (-92.29) MeV. At finite density, the magnitude of scalar fields increase with the increase in the temperature of the medium. For example, at $\rho_N$=$\rho_0$ ($4\rho_0$), $\eta$=0 and $eB$=4${{m^2_{\pi}}}$ the values of $\sigma$ fields are -58.11 (-28.57), -62.69 (-31.65) and -63.98 (-33.68) MeV for $T$=0, 100 and 150 MeV, respectively and for $\zeta$ field, these values are -96.91 (-92.63), -97.97 (-92.93) and -98.25 (-93.15) MeV.

Considering the effects of isospin asymmetry in  subplots (c),(d),(g) and (h), it is observed that at $\eta$=0.5 and $\rho_N$=4$\rho_0$, the magnitude of $\sigma$ and $\zeta$ field decreases (same as of symmetric matter case) with respect to $eB$ for temperature $T$=100 and 150 MeV, whereas it increase for low  temperature such as $T$=0 and 50 MeV.
  This is due to the fact that in highly asymmetric matter, due to large number of neutrons, a significant difference in  $\rho^{s}_n$ and  $\rho^{s}_p$ is observed which results in opposite behaviour of fields in low temperature.   Also, for the same $\eta$,  the temperature effect on $\sigma$ and $\zeta$ field  become very less for high density with respect to magnetic field. The above behaviour is due to the competing effects between scalar densities $\rho^{s}_n$ and $\rho^{s}_p$ of neutron and proton  as these densities have the temperature and magnetic field dependence \cite{Kumar2018,Kumar2010}.

In Fig.\ref{fdeltachi}, the scalar  fields $\delta$ and $\chi$ are plotted as a function of magnetic field. It is observed that in subplot (a)-(d), the magnitude of $\delta$ field increases with respect to magnetic field and it varies more  appreciably in high density as compared to low density. For $\eta$=0, the non-zero value of $\delta$ field is observed due to the inequality $\rho^{s}_n$ $\neq$ $\rho^{s}_p$, as for finite magnetic field the Lorentz force effects comes into picture\cite{Kumar2018}, which acts on neutron and proton differently. Further, if we move towards higher asymmetry, the temperature and magnetic field effects on  $\delta$ field are more prominent. For example, at $\rho_N$=4$\rho_0$, $eB$=4${{m^2_{\pi}}}$ and $\eta$=0 (0.5) the values of $\delta$ field are 0.70 (-4.24), 0.62 (-3.73) and 0.52 (-2.39) MeV for $T$=0, 100 and 150 MeV respectively and for  $eB$=7${{m^2_{\pi}}}$, these values changes to 1.82 (-4.24), 1.65 (-2.62) and 1.58 (-0.26) MeV. In subplot (e)-(h), the glueball field $\chi$ is plotted and it is noticed that for $\eta$=0, the magnitude of the dilaton field decrease with the increase in magnetic field. Here, the observed decrement is more in high density regime as compared to low density regime. 
 Furthermore, in pure neutron matter, the trend of $\chi$ field is exactly opposite as compared to symmetric medium for low temperature. For example, at $\eta$=0.5, $\rho_N$=$4\rho_0$ and $eB$=$4{{m^2_{\pi}}}$ (7${{m^2_{\pi}}}$), the values of $\chi$ fields are observed as 403.6 (404.3), 404 (403.3) and 403.2 (400.4) MeV for $T$=0, 100 and 150 MeV respectively. Whereas for $\eta$=0 and same parameters, the values are   401.1(398.1),  402.6(400) and  403.5(400.6) MeV. This observed crossover is due to the fact that the $\chi$ field is solved along with $\sigma$ and $\zeta$ fields in coupled equations of motions, and which as discussed earlier, contributes from the magnetic field induced scalar densities of neutron and proton.

\begin{figure}
\includegraphics[width=16cm,height=21cm]{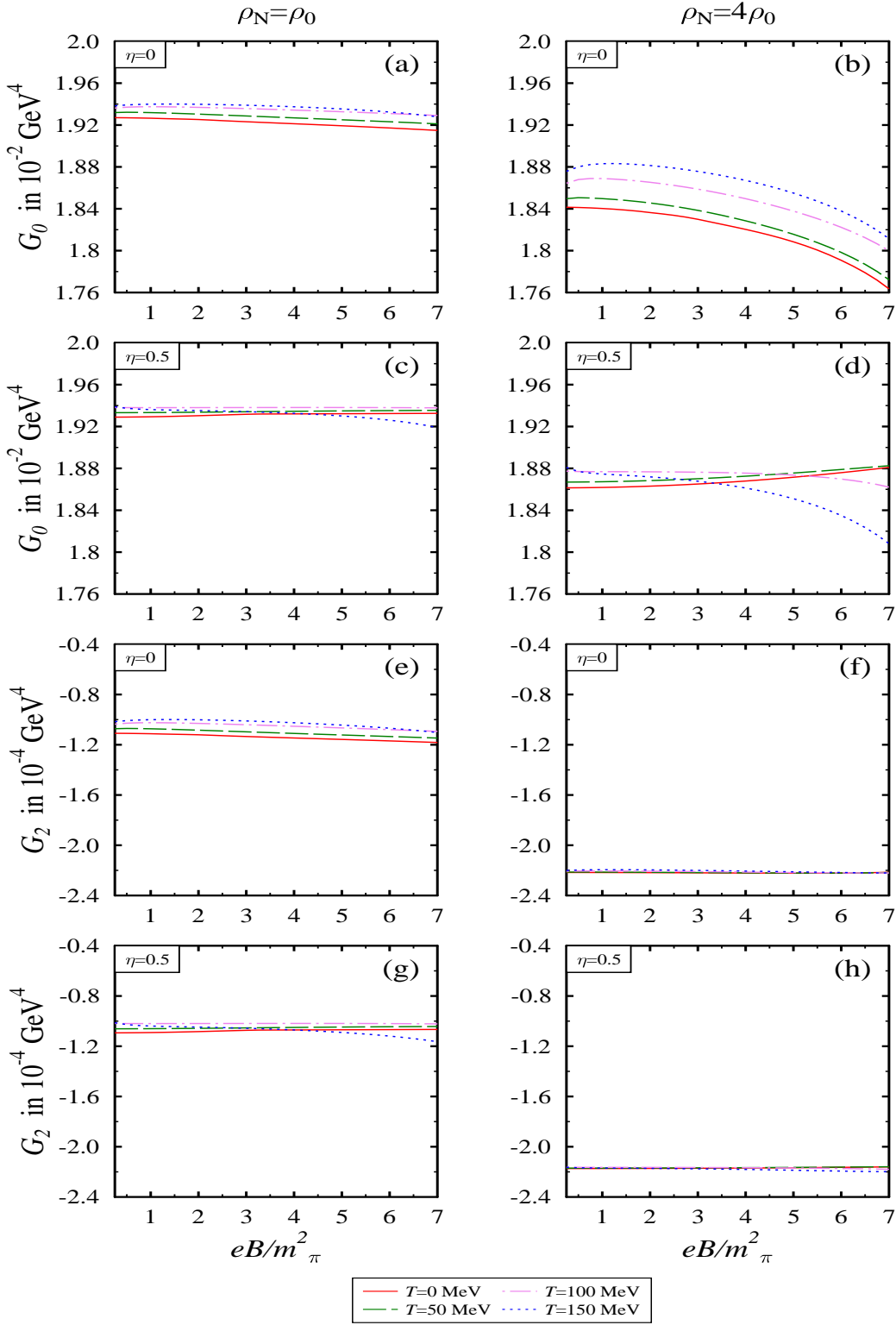}
\caption{(Color online) The scalar gluon condensate $G_0$ and twist-2 gluon operator $G_2$, describing the trace and nontrace part of energy momentum tensor respectively plotted as a function of magnetic field $eB/{{m^2_{\pi}}}$, for different values of temperature $T$, nucleon density $\rho_N$ and isospin asymmetry parameter, $\eta$.}
\label{fG0G2}
\end{figure}

\subsection{In-medium gluon condensates and mass-shift of charmonia and bottomonia} 

In Fig.\ref{fG0G2}, We have plotted scalar and twist-2 gluon operator condensates with magnetic field for different values of nucleon density, temperature and isospin asymmetry . As can be seen from Eqs.(\ref{chiglum}) and (\ref{g2approx}), the scalar and tensorial condensate depends on the scalar fields and both of them strongly depends on $\chi$ field due to its fourth power dependence. Therefore, the behaviour of these condensates reflects the same functional dependence on magnetic field and other parameters as $\chi$ field does. In the present investigation, we get non-zero value of twist-2 gluon operator $G_2$, as magnetic field  breaks down the Lorentz symmetry by affecting the neutron and proton separately \cite{Klingl1999,Kumar2018}. In addition, in high density regime the variation of $G_2$ is almost negligible due to the presence of second term in  Eq.(\ref{g2approx}). It is observed that at high density  the values of $G_0$ and $G_2$ decrease with the increase in magnetic field for symmetric nuclear matter, while it increase for $T$=0 and 50 MeV for $\eta$=0.5.

\begin{table}[h]
\centering
\begin{tabular}{|l|l|l|l|l|l|l|l|l|l|l|l|l|l|}
\hline
\multirow{3}{*}{}  & \multirow{3}{*}{} & \multicolumn{4}{c|}{$\eta=0$}                          & \multicolumn{4}{c|}{$\eta=0.3$}                         & \multicolumn{4}{c|}{$\eta=0.5$}                         \\ \cline{3-14} 
                   &          $\Delta m_Q$          & \multicolumn{2}{c|}{$T$=0} & \multicolumn{2}{c|}{$T$=100} & \multicolumn{2}{c|}{$T$=0} & \multicolumn{2}{c|}{$T$=100} & \multicolumn{2}{c|}{$T$=0} & \multicolumn{2}{c|}{$T$=100} \\ \cline{3-14} 
                 &  &$\rho_0$ &$4\rho_0$ &$\rho_0$  &$4\rho_0$ &$\rho_0$ &$4\rho_0$ &$\rho_0$  &$4\rho_0$ &$\rho_0$ &$4\rho_0$ &$\rho_0$ &$4\rho_0$ \\ \hline
\multirow{2}{*}{${\chi_c}_0$} &        $\Delta m_{Q_{30}}$            &-3.09 &-12.37 &-1.92 & -10.04 & -2.63 &-11.57 &-1.77 &-9.26&-2.30 &-9.47 &-1.69 &-8.58 \\  \cline{2-14}
                   &         $\Delta m_{Q_{70}}$ &-3.83 & -17.83 &-2.54 &-14.85&-2.81 &-13.67 & -1.94& -11.14 & -2.22 &-8.21 &-1.72& -9.72 \\  \hline
\multirow{2}{*}{${\chi_c}_1$} &     $\Delta m_{Q_{30}}$                           &-4.15&-16.29&-2.62&-13.34&-3.55&-15.30&-2.41&-12.32&
-3.09&-12.60&-2.32&-11.43\\  \cline{2-14}
                   &        $\Delta m_{Q_{70}}$                        &-5.13&-23.44&-3.43&-19.53&-3.79&-17.97&-2.64&-14.77
&-3.01&-10.95&-2.36&-12.92\\ 

\hline
\end{tabular}
\caption{In the above table,  we tabulate the  effect of magnetic field on the mass-shifts of  ${\chi_c}_0$ and ${\chi_c}_1$ mesons from Borel sum rule. Here, $\Delta m_{Q_{30}}$ represents mass-shift between $eB$=3${{m^2_{\pi}}}$ and B=0 (similar for $\Delta m_{Q_{70}}$). In addition, $\rho_N$ and temperature $T$  are given in units  of fm$^{-3}$ and MeV respectively. }
\label{table_charm}
\end{table}

\begin{figure}
\includegraphics[width=16cm,height=8cm]{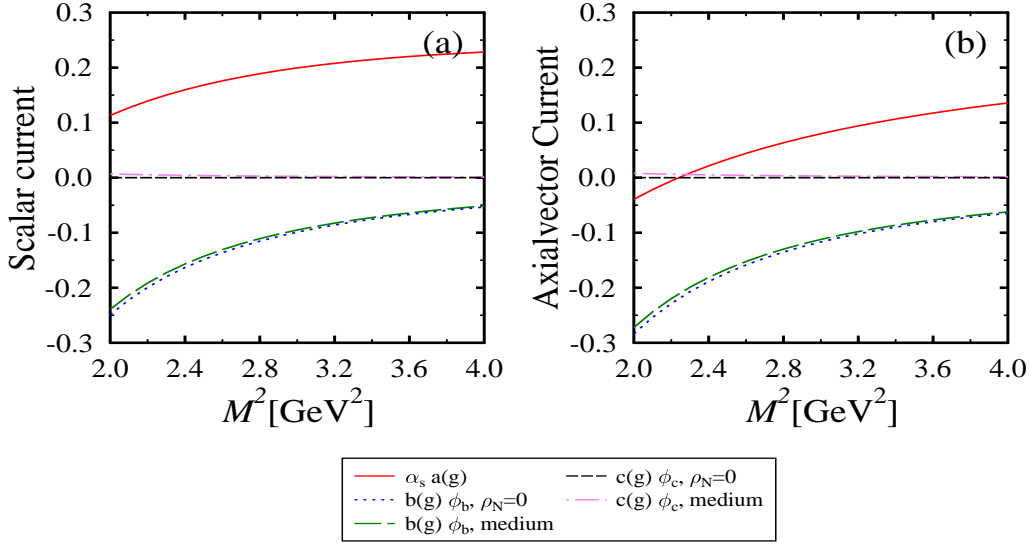}
 \caption{(Color online)The OPE coefficients(charm) for scalar and axialvector current  plotted as a function of Borel parameter $M^2$, for $eB=5{{m^2_{\pi}}}$, $T$=100 MeV, nucleon density $\rho_N$=$4\rho_0$ and isospin asymmetry parameter, $\eta$=0.3.}
\label{fopesa}
\end{figure}

\begin{figure}
\includegraphics[width=14cm,height=14cm]{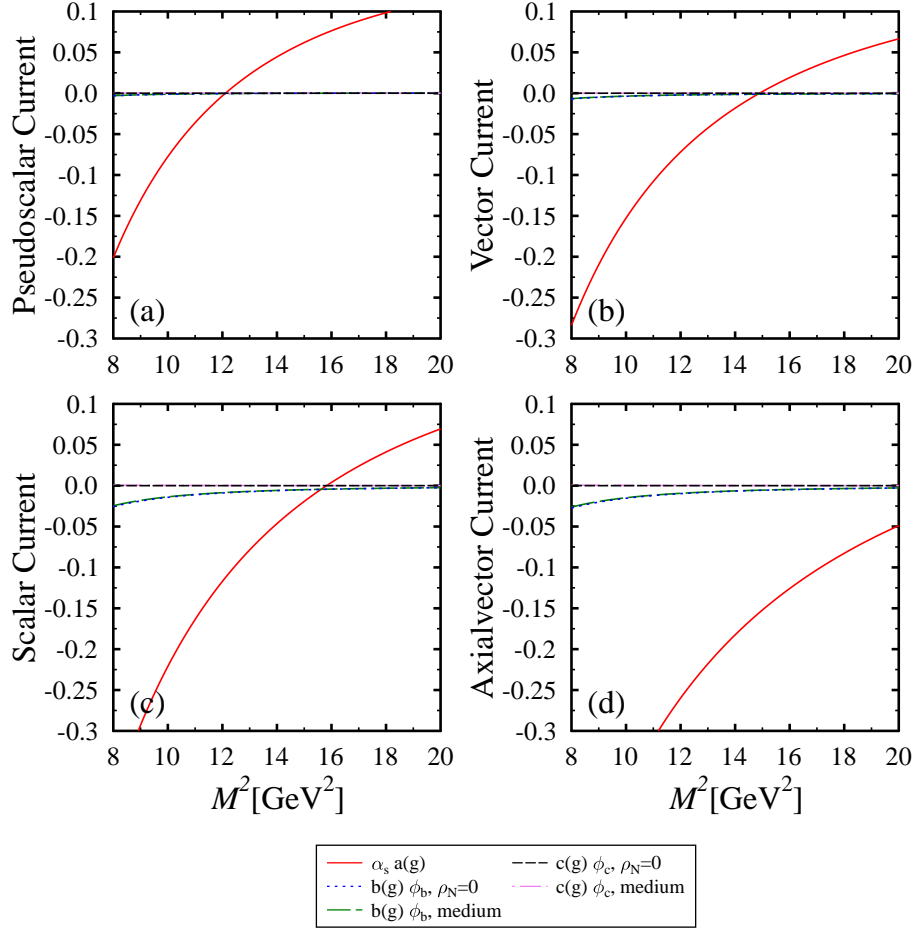}
 \caption{(Color online)The OPE coefficients(bottom) for pseudoscalar, vector, scalar and axialvector current  plotted as a function of Borel parameter $M^2$, for $eB=5{{m^2_{\pi}}}$, $T$=100 MeV, nucleon density $\rho_N$=$4\rho_0$ and isospin asymmetry parameter, $\eta$=0.3.}
\label{fopepvsab}
\end{figure}

\begin{figure}
\includegraphics[width=16cm,height=21cm]{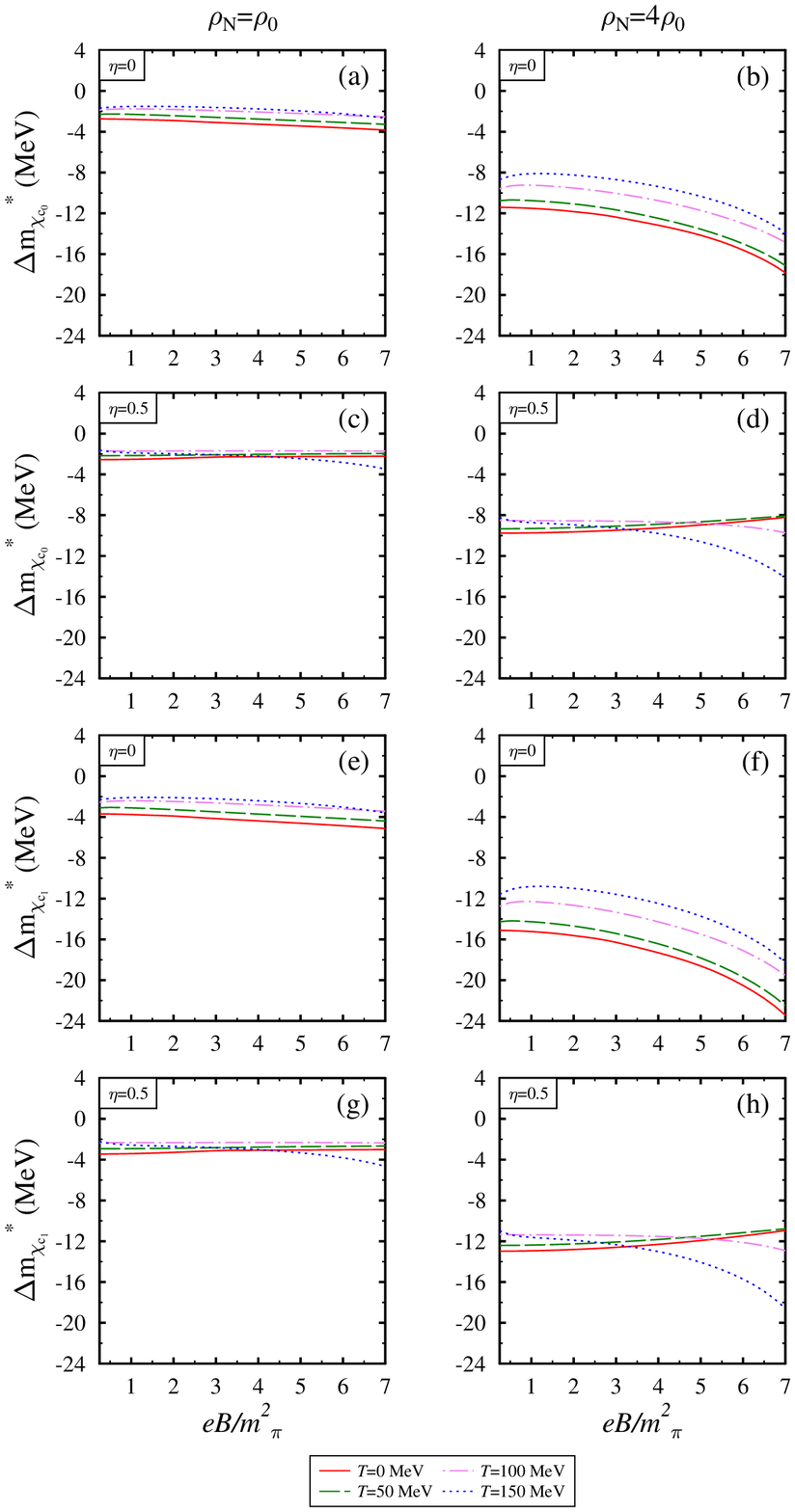}
 \caption{(Color online)The in-medium mass-shift of ${\chi_c}_0$ and  ${\chi_c}_1$ meson calculated from Borel sum rule plotted as a function of magnetic field $eB/{{m^2_{\pi}}}$, for different values of temperature $T$, nucleon density $\rho_N$ and isospin asymmetry parameter, $\eta$.}
\label{fcc0cc1}
\end{figure}

\begin{figure}
\includegraphics[width=16cm,height=8cm]{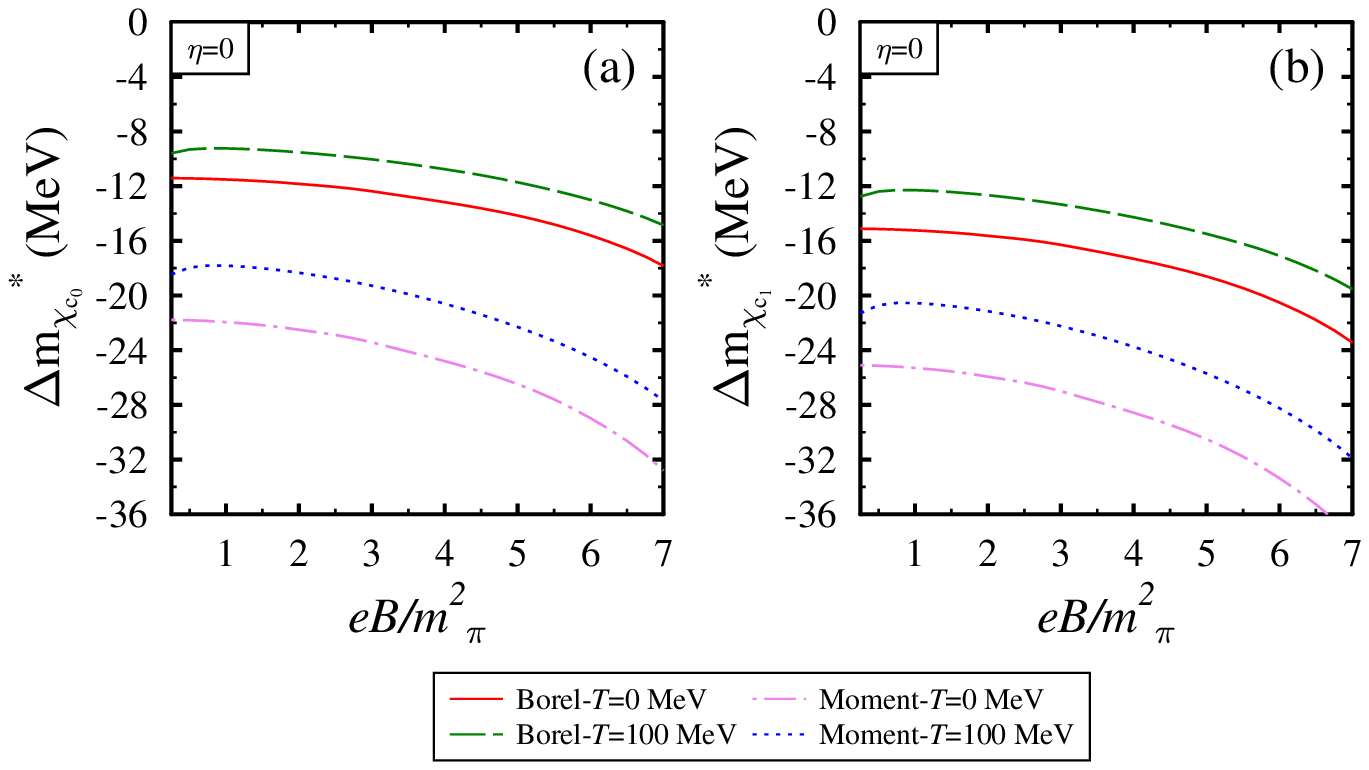}
 \caption{(Color online)The in-medium mass-shift of ${\chi_c}_0$ and  ${\chi_c}_1$ meson plotted as a function of magnetic field $eB/{{m^2_{\pi}}}$ by using different sum rules, for $T$=0 and 100 MeV, nucleon density $\rho_N$=$4\rho_0$ and isospin asymmetry parameter, $\eta$=0.}
\label{fmassshiftscbm}
\end{figure}

These condensates are used as  input in $\phi_b$ and $\phi_c$  to calculate the effective mass and  hence, in-medium mass-shift of quarkonia from Borel and moment sum rule. In the former case, the threshold values $s_0$ is taken as to get the stability of  the plot between effective mass  with respect to Borel parameter $M^2$ and in this article, we have taken its value to be 12.25 and 114.91 GeV$^2$ for charmonia and bottomnia, respectively\cite{Morita2010a}. To understand the implications from the OPE side, in Figs.(\ref{fopesa}) and (\ref{fopepvsab}), we have plotted the OPE coefficients for Borel sum rule with respect to Borel parameter. In Fig.\ref{fopesa}, for $P$-wave charmonia, we observed for a particular value of medium parameters($eB,T,\rho_N$ and $\eta$), the expansion coefficients of charmonia varies appreciably with Borel parameter. However, in   Fig.\ref{fopepvsab}, for same parameters, the bottomonia's  expansion coefficients such as $b(g)\phi_b$ and $c(g)\phi_c$ are suppressed due to the term $m_q^2$ present in the expression of $\phi_b$ and $\phi_c$, as the mass of bottom quark is very large. From both  sum rules, it is observed that the effective mass of the quarkonia, decreases with the increase in magnetic field, temperature, nucleon density and isospin asymmetry.  The observed mass-shift of axial vector meson ${\chi_c}_1$ is more than the scalar meson ${\chi_c}_0$. In Fig.\ref{fcc0cc1},  we have plotted the in-medium mass-shift(Borel sum rule) of ${\chi_c}_0$ and ${\chi_c}_1$ mesons as a function of magnetic field for different values of isospin asymmetry, temperature and nucleon density. In this figure, for $\eta$=0, we noticed that the in-medium negative mass-shift increases more with the increase in magnetic field. Considering the temperature effects, the magnitude of mass-shift is more at zero temperature for both nucleon densities. It decreases with the increase in temperature but follows the same functional dependence for magnetic field.  
For highly asymmetric nuclear matter, at $\rho_N$=4$\rho_0$, the magnitude of in-medium mass-shift decreases with the increase in magnetic field $eB$ for $T$=0 and 50 MeV, whereas for $T$=100 and 150 MeV, the magnitude of mass-shift increases with the increase in magnetic field. The values of mass-shift of $P$-wave charmonia calculated from Borel sum rule at $eB$=3${{m^2_{\pi}}}$ and 7${{m^2_{\pi}}}$ are tabulated in \cref{table_charm}. 
  The crossover is a reflection of values of $\phi_b$ and $\phi_c$, which are directly proportional to gluon condensates $G_0$ and $G_2$, respectively. In Fig.\ref{fmassshiftscbm}, we have compared  the mass-shift of $P$-wave charmonia obtained from Borel sum rule as well as moment sum rule as a function of magnetic field and for fixed value of nucleon density, temperature and isospin asymmetry parameter.  For moment sum rule, following our previous work for $J/\psi$ and $\eta_c$ mesons \cite{Kumar2018}, in this article we work with $\xi=2.5$, since the Wilson coefficients for heavy quarkonia are larger than the $S$-wave \cite{Song2009}. We found that the  magnitude of observed mass-shift of quarkonia is less for Borel sum rule as compared to moment sum rule. For example, in moment sum rule, the values of ${\chi_c}_0$ mass-shift at $\rho_N$=$4\rho_0$ and $eB$=3${{m^2_{\pi}}}$(7${{m^2_{\pi}}}$) with respect to vacuum  are   -23.42(-32.79) and -19.28(-27.68) MeV for  $T$=0 and 100 MeV, respectively and for the  ${\chi_c}_1$ meson  mass-shift for same parameters  are   -26.99(-37.74) and -22.24(-31.89) MeV. The difference in the mass-shift from two sum rules is because in Borel sum rule for better perturbative expansion, the limit on the dispersion relation(see Eq.(\ref{Borel_moment})) goes to more deeper euclidean region.    In Ref.\cite{Song2009}, the author have reported the  mass-shift of $\chi_{c_0}$ and $\chi_{c_1}$  as a function of $T/T_c$ in zero magnetic field and  observed that the effective masses of these mesons decrease with the increase in $T/T_c$. 
  
%

\begin{figure}
\includegraphics[width=16cm,height=21cm]{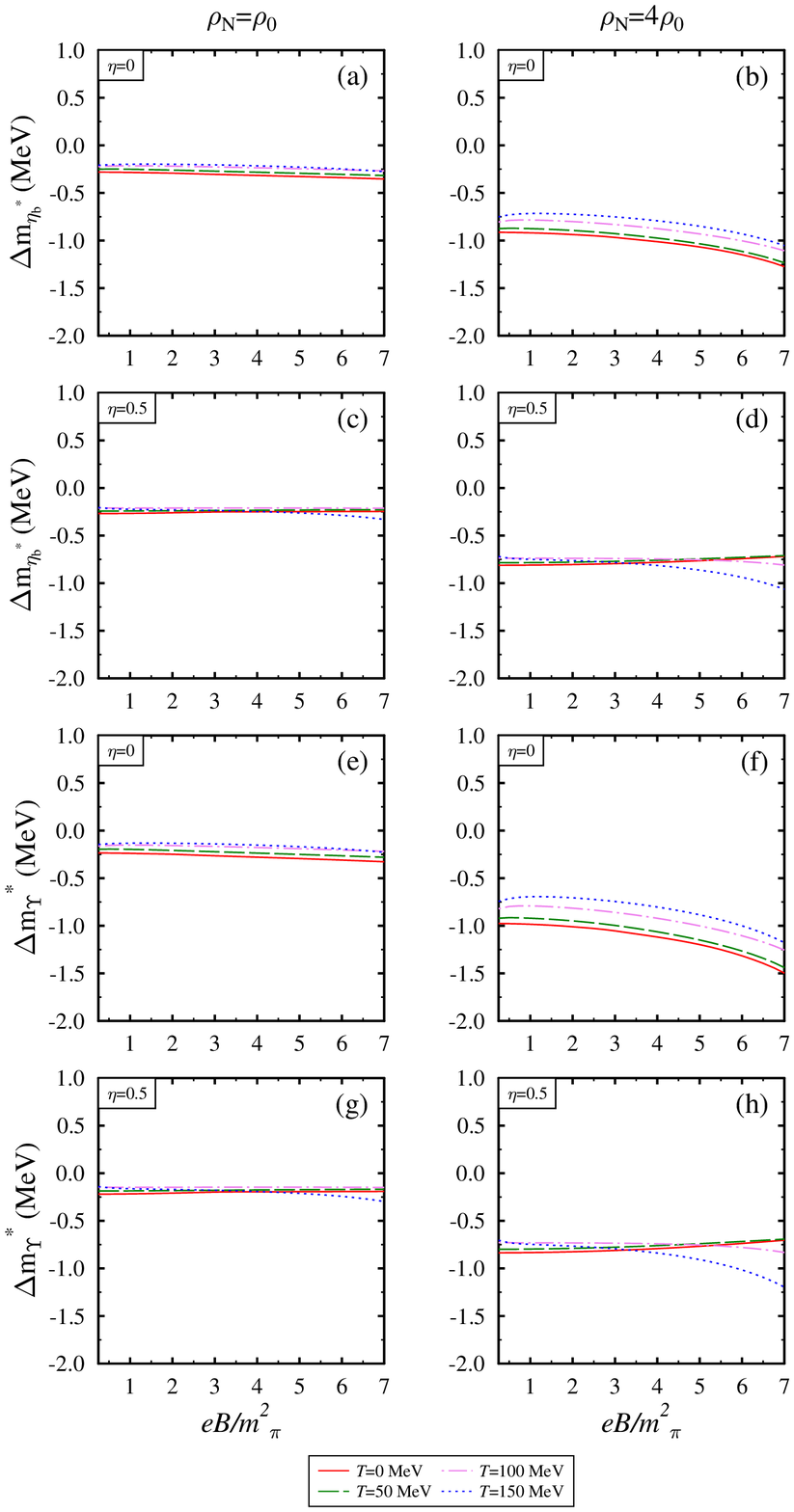}
 \caption{(Color online)The in-medium mass-shift of $\eta_b$ and  $\Upsilon$ meson calculated from Borel sum rule plotted as a function of magnetic field $eB/{{m^2_{\pi}}}$, for different values of temperature $T$, nucleon density $\rho_N$ and isospin asymmetry parameter, $\eta$.}
\label{fupeta}
\end{figure}

\begin{figure}
\includegraphics[width=16cm,height=21cm]{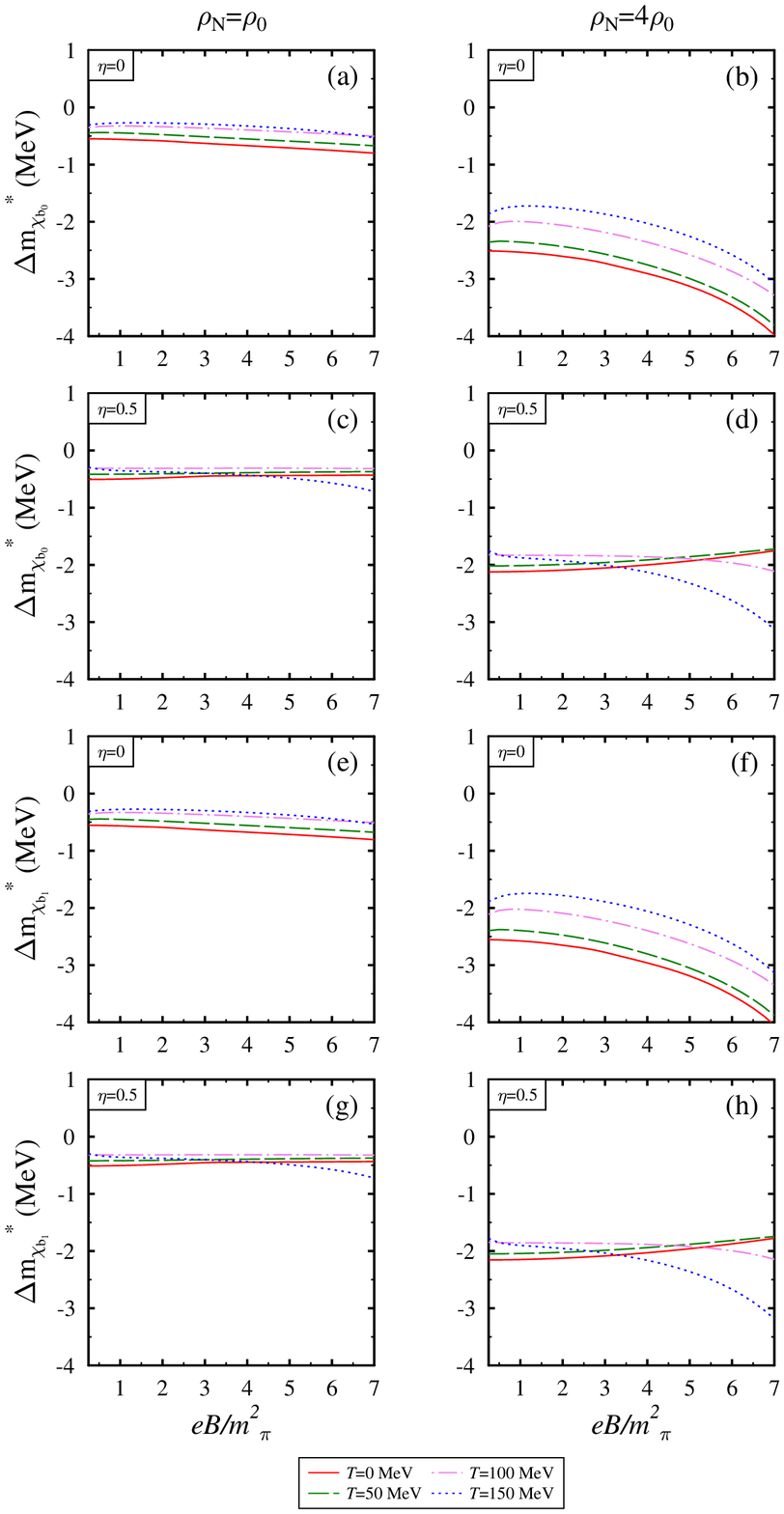}
 \caption{(Color online)The in-medium mass-shift of ${\chi_b}_0$ and  ${\chi_b}_1$ meson calculated from Borel sum rule plotted as a function of magnetic field $eB/{{m^2_{\pi}}}$, for different values of temperature $T$, nucleon density $\rho_N$ and isospin asymmetry parameter, $\eta$.}
\label{fcb0cb1}
\end{figure}

  In Figs.\ref{fupeta} and \ref{fcb0cb1}, we have plotted the in-medium magnetic field induced mass shift of bottomonia with respect to magnetic field  using Borel sum rule. In Fig.\ref{fupeta}, the mass-shift of pseudoscalar  meson $\eta_b$ and vector meson $\Upsilon$ is plotted, whereas in Fig.\ref{fcb0cb1}, scalar meson $\chi_{b_0}$ and pseudovector meson $\chi_{b_1}$ is plotted. The mass-shift of bottomonia follows the same trend as for charmonia for different medium parameters. However, we found that the effect of medium characteristics  on the mass of bottomonia is very less as compared to charmonia's mass. This is due to the presence of square mass term in  Eqs.(\ref{phib}) and (\ref{phic}) as discussed earlier.  In the bottom sector, we can see that the magnitude of mass-shift increase with the increase in nuclear density. The induced bottomonia mass-shifts are listed in \cref{table_bottom}. If we compare this Borel sum rule result with moment sum rule results, we found that the obtained mass-shift from Borel sum rule little different than the moment sum rule. For example, in moment sum rule, the values of $\eta_b$ mass-shift at $\rho_N$=$4\rho_0$ and $eB$=3${{m^2_{\pi}}}$(7${{m^2_{\pi}}}$) with respect to vacuum  are   -0.56(-0.71) and -0.49(-0.63) MeV for  $T$=0 and 100 MeV, respectively and for the  $\Upsilon$ meson  mass-shift for same parameters  are   -0.67(-0.95) and -0.54(-0.79) MeV. Similarly, the values of ${\chi_b}_0$ mass-shift at $\rho_N$=$4\rho_0$ and $eB$=3${{m^2_{\pi}}}$(7${{m^2_{\pi}}}$) with respect to vacuum  are   -3.11(-4.46) and -2.51(-3.72) MeV for  $T$=0 and 100 MeV, respectively and for the  ${\chi_b}_1$ meson  mass-shift for same parameters  are   -2.05(-3.67) and -2.06(-3.06) MeV.  In Ref.\cite{Jahanb2018}, author have calculated the in-medium mass-shift of Upsilon states in the presence of magnetic field. In this article, for $\eta$=0, at  $\rho_N$=$\rho_0$($4\rho_0$), the mass-shift of $\Upsilon$(1s) are calculated as -0.71(-2.57) and  -0.71(-2.46) MeV for magnetic fields 4${{m^2_{\pi}}}$ and  8${{m^2_{\pi}}}$ respectively at zero temperature, which are consistent with our results. The QCDSR along with Maximum Entropy Method(MEM)has been applied  to study the thermal effects on bottomonia  at zero magnetic field\cite{Suzuki2013}. In this article, the authors have studied the dissociation of excited state of bottomonia at lower temperature than of ground state.  In the next section, we will now conclude our present work.

\begin{table}[h]
\centering
\begin{tabular}{|l|l|l|l|l|l|l|l|l|l|l|l|l|l|}
\hline
\multirow{3}{*}{}  & \multirow{3}{*}{} & \multicolumn{4}{c|}{$\eta=0$}                          & \multicolumn{4}{c|}{$\eta=0.3$}                         & \multicolumn{4}{c|}{$\eta=0.5$}                         \\ \cline{3-14} 
                   &          $\Delta m_Q$          & \multicolumn{2}{c|}{$T$=0} & \multicolumn{2}{c|}{$T$=100} & \multicolumn{2}{c|}{$T$=0} & \multicolumn{2}{c|}{$T$=100} & \multicolumn{2}{c|}{$T$=0} & \multicolumn{2}{c|}{$T$=100} \\ \cline{3-14} 
                 &  &$\rho_0$ &$4\rho_0$ &$\rho_0$  &$4\rho_0$ &$\rho_0$ &$4\rho_0$ &$\rho_0$  &$4\rho_0$ &$\rho_0$ &$4\rho_0$ &$\rho_0$ &$4\rho_0$ \\ \hline
\multirow{2}{*}{$\eta_b$} &        $\Delta m_{Q_{30}}$            &-0.30 &-0.96 &-0.22 & -0.83 & -0.27 &-0.92 &-0.21 &-0.78&-0.25 &-0.79 &-0.21 &-0.74 \\  \cline{2-14}
                   &         $\Delta m_{Q_{70}}$ &-0.35 & -1.27 &-0.26 &-1.10 &-0.28 &-1.04 & -0.22& -0.89 & -0.24 &-0.71 &-0.21& -0.81 \\  \hline
\multirow{2}{*}{$\Upsilon$} &     $\Delta m_{Q_{30}}$                           &-0.26&-1.05&-0.16&-0.85&-0.22&-0.98&-0.15&-0.79&
-0.19&-0.81&-0.14&-0.73\\  \cline{2-14}
                   &        $\Delta m_{Q_{70}}$                        &-0.32&-1.49&-0.21&-1.25&-0.24&-1.15&-0.16&-0.95
&-0.19&-0.70&-0.14&-0.83\\  \hline
\multirow{2}{*}{${\chi_b}_0$} &     $\Delta m_{Q_{30}}$                           &-0.62&-2.72&-0.36&-2.18&-0.52&-2.54&-0.32&-2.00&
-0.44&-2.00&-0.31&-1.84\\  \cline{2-14}
                   &        $\Delta m_{Q_{70}}$                        &-0.79&-3.97&-0.50&-3.28&-0.56&-3.02&-0.36&-2.44
&-0.43&-1.75&-0.31&-2.12\\  \hline
\multirow{2}{*}{${\chi_b}_1$} &     $\Delta m_{Q_{30}}$                           &-0.63&-2.77&-0.36&-2.22&-0.52&-2.90&-0.33&-2.03&
-0.45&-2.08&-0.31&-0.86\\  \cline{2-14}
                   &        $\Delta m_{Q_{70}}$                        &-0.80&-4.04&-0.50&-3.35&-0.50&-3.07&-0.37&-2.48
&-0.43&-0.77&-0.32&-2.14\\ 

\hline
\end{tabular}
\caption{In the above table,  we tabulate the  effect of magnetic field on the mass-shifts of $\eta_b$, $\Upsilon$,  ${\chi_b}_0$ and ${\chi_b}_1$ mesons from Borel sum rule. Here, $\Delta m_{Q_{30}}$ represents mass-shift between $eB$=3${{m^2_{\pi}}}$ and B=0 (similar for $\Delta m_{Q_{70}}$). In addition, $\rho_N$ and temperature $T$  are given in units  of fm$^{-3}$ and MeV respectively. }
\label{table_bottom}
\end{table}

\section{Conclusions}
\label{sec:5}

In the present investigation, we calculated the mass-shift of $P$-wave charmonium and $S$ and $P$-wave bottomonium. It is noticed that the  magnetic field induced mass-shift of $P$-wave charmonia are more prominent than $S$-wave charmonia\cite{Kumar2018}, whereas the change in the  mass-shift is very less for bottomonia. To calculate the effective mass, we have used the unified approach of chiral $SU(3)$ model and QCDSR. In the former, the medium effects are calculated in terms of $\sigma$, $\zeta$, $\delta$ and $\chi$ mesonic fields which are used further to calculate the in-medium scalar  and tensorial gluon condensates. Using these gluon condensates as  input in the QCDSR, the mass-shift of ${{\chi_c}_0}$, ${{\chi_c}_1}$, $\eta_b$, $\Upsilon$, ${{\chi_b}_0}$ and ${{\chi_b}_1}$
mesons have been calculated. The temperature and magnetic field effects are introduced through the scalar density $\rho^{s}_i$, and vector density $\rho^{v}_i$, of the $i^{th}$ nucleons \cite{Kumar2018}. Also, we have introduced the effect of isospin asymmetry by the incorporation of 
$\delta$ and $\rho$ fields which are solved along with other mesonic fields through the coupled equations of motion. Within QCDSR, we use two different sum rules approaches ,$i.e.$, Borel sum rule and moment sum rule. In this article, we have observed that from both sum rules, the effective mass of these mesons decreases with the increase in magnetic field and nucleon density. The magnitude of in-medium mass-shift increases with the increase in magnetic field for all temperatures except for pure neutron matter, where it increases for high temperature but decreases for low temperature.  Also, the decrease of in-medium mass-shift of above mesons is more in case of high-density regime than low density. However, the magnitude of mass-shift observed from both the sum rules is different. We observe less magnitude of mass-shift in Borel sum rule than in moment sum rule.  The calculated mass-shift can be used to study the in-medium decay of higher quarkonium states \cite{Song2009,Friman2002} to lower quarkonium states for a better understanding of $J/\psi$ suppression.  This decay width can also be used to calculate the experimental observables like in-medium cross-section \cite{Khachatryan2017} in asymmetric heavy ion collision experiments.

\begin{center}
 \section*{Acknowledgment}
\end{center}

One of the author, Rajesh Kumar sincerely acknowledge the support towards this work from Ministry of Science and Human Resources Development (MHRD), Government of India via Institute fellowship under National Institute of Technology Jalandhar.

\end{document}